\documentclass{aastex}
\usepackage{emulateapj5}

\shorttitle{Three Wide Separation Companions} \shortauthors{Wilson
et al.}

\begin{document}
\submitted{AJ, accepted 03 Jul 2001}
\title{Three Wide Separation L-dwarf Companions
from the Two Micron All Sky Survey: \linebreak Gl 337C, Gl 618.1B, and HD
89744B\altaffilmark{1}}

\author{J.C. Wilson\altaffilmark{2}, J. Davy Kirkpatrick\altaffilmark{3},
J.E. Gizis\altaffilmark{4}, M.F.
Skrutskie\altaffilmark{5},\linebreak  D.G. Monet\altaffilmark{6}
and J.R. Houck\altaffilmark{7}}

\altaffiltext{1}{Portions of the data presented here were
obtained at the W. M. Keck Observatory, which is operated as a
scientific partnership among the California Institute of
Technology, the University of California, and the National
Aeronautics and Space Administration. The Observatory was made
possible by the generous financial support of the W. M. Keck
Foundation.  Observations at the Palomar Observatory were made as
part of a continuing collaboration between the California
Institute of Technology and Cornell University. The 60-inch
telescope at Palomar Mountain is jointly owned by the California
Institute of Technology and the Carnegie Institution of
Washington.}
\altaffiltext{2}{Space Sciences Building, Cornell
University,
    Ithaca, NY 14853; jcw14@cornell.edu}
\altaffiltext{3}{Infrared Processing and Analysis Center, M/S
    100-22, California Institute of Technology, Pasadena, CA 91125;
    davy@ipac.caltech.edu}
\altaffiltext{4}{Infrared Processing and Analysis Center, M/S
    100-22, California Institute of Technology, Pasadena, CA
    91125; gizis@whitesands.ipac.caltech.edu}
\altaffiltext{5}{Department of Astronomy, University of
    Massachusetts, Amherst, MA 01003; skrutski@north.astro.umass.edu}
\altaffiltext{6}{U.S. Naval Observatory, P.O. Box 1149, Flagstaff,
    AZ 86002; dgm@nofs.navy.mil}
\altaffiltext{7}{Space Sciences Building, Cornell University,
    Ithaca, NY 14853; jrh13@cornell.edu}

\begin{abstract}
We present two confirmed wide separation L-dwarf common proper
motion companions to nearby stars and one candidate identified
from the Two Micron All Sky Survey. Spectral types from optical
spectroscopy are L0 V, L2.5 V, and L8 V. Near-infrared low
resolution spectra of the companions are provided as well as a
grid of known objects spanning M6 V -- T dwarfs to support
spectral type assignment for these and future L-dwarfs in the
\textit{z'JHK} bands. Using published measurements, we estimate
ages of the companions from physical properties of the primaries.
These crude ages allow us to estimate companion masses using
theoretical low-mass star and brown dwarf evolutionary models. The
new L-dwarfs in this paper bring the number of known wide-binary
($\Delta \geq 100$ AU) L-dwarf companions of nearby stars to nine.
One of the L-dwarfs is a wide separation companion to the F7 IV-V
$+$ extrasolar planet system HD89744Ab.
\end{abstract}


\section{Introduction}
The study of binary and multiple systems in the local neighborhood
is important for several reasons.  It is desirable to have an
accurate accounting of stellar systems in the local neighborhood.
Improving the statistics of binary primary masses, system mass
fractions, orbital separations, and eccentricities are crucial for
refining stellar and planetary formation theories. Also, studying
wide separation systems in particular gives insight into their
dynamic stability with time as a function of orbital separation
and mass fraction (and hence binding energy) against disruptive
events such as close approaches to giant molecular clouds, stars,
and the gravitational potential of the inner Galaxy.

In this paper we present three wide separation L-dwarf companions
to nearby stars. One companion has an apparent separation
$(\Delta)$ of 100 AU $< \Delta <1000$ AU, and two have
$\Delta>1000$ AU. These discoveries, along with similar companions
previously reported, highlight the success in searching new
portions of binary system phase space (i.e. wide separations) by
surveys such as Two Micron All Sky Survey (2MASS; \citet{skr97}),
the Deep Near-Infrared Survey (DENIS; \citet{epc97}), and the
Sloan Digital Sky Survey (SDSS; \citet{yor00}).  The surveys'
sensitivities and large areal coverages allow for very low-mass
(VLM) stars and brown dwarf companions to be found at separations
$\Delta>100$ AU.\footnote{Unless otherwise stated we use the term
`separation', ($\Delta$), to mean `apparent separation.'} The new
detections will in turn improve the statistics of systems with
mass ratios ($q$) far from unity.\footnote{We define $q \equiv
m_{sec}/m_{pri}$, and use the term `low mass ratio' to mean $q$
near zero.}  For instance, the low-mass companions already
discovered show that the `brown dwarf desert' at close separation
($<1$ \% of FGKM main sequence primaries have brown dwarf
companions at $\Delta<3$ AU; \citet{mar00}) may not extend to wide
separations \citep{gi01a}. The added statistics can also be used
to test the mass ratio distribution results of \citet{duq91}:
their distribution peaks at $q\sim 0.3$ and is then either flat or
decreases for $q<0.2$. \citet{duq91}, however, used the radial
velocity technique for searching for unknown companions, a
technique poorly suited for discovering $q<0.1$ systems at wide
separation.

The study of VLM stars and brown dwarfs as singular objects also
benefits from the discovery of these objects as companions in
binary systems.  Because brown dwarfs continually cool with time
since they cannot sustain hydrogen fusion, and VLM stars near the
main sequence limit can take a Hubble time to come to equilibrium,
degeneracies occur in the HR diagram for these objects.  For
example, a field L1 V can either be a very old and VLM star or a
less massive, young brown dwarf that is slowly cooling to later
spectral types.  Binary systems can be used to break this
degeneracy and provide examples of L-dwarfs with estimated masses
as follows: assuming that the primary and low-mass companion in a
system are coeval, age estimates from the primary can be adopted
for the secondary. The ages along with inferred effective
temperatures can then be used to determine crude mass estimates
from theoretical evolutionary curves. Metallicity and distance
information is also acquired for L-dwarfs in binary systems from
known properties of the primaries.

There are six previously discovered L-dwarfs that are wide
separation ($\Delta \geq 100$ AU) companions to nearby stars (see
\citet{kir01}, Table 7a, for a summary), including three from
2MASS.  The latter were L-dwarf candidates found to be
serendipitously located close on the sky to known nearby stars. In
this paper we present the first results from a \it{dedicated} \rm
2MASS search for L-dwarf candidates in close proximity on the sky
to known nearby stars.  The search area around any given candidate
primary is an annulus with inner radius defined by the object
brightness and the capability of the 2MASS point source extraction
algorithm to identify faint objects adjacent to bright sources.
The outer radius is defined by the distance to the primary and the
maximum companion separation plausible from dynamical binding
considerations.  The candidate primaries are mainly drawn from the
Catalogue of Nearby Stars \citep{gli91} and other nearby objects
discovered more recently.  A future paper will discuss in more
detail the search criteria and success rate. In this paper we
present two confirmed (spectroscopic and common proper-motion)
companions discovered while searching the vicinity of two `Gliese'
objects: 2MASSW J0912145+145940 (L8 V companion to a G8 V$+$K1 V
spectroscopic binary) and 2MASSW J1620261-041631 (L2.5 V companion
to an M0 V).

We also present the spectroscopically (but not common
proper-motion) confirmed candidate companion 2MASSI
J1022148+411426 (L0 V).\footnote{Objects with designations 2MASSI
are part of the Second Incremental Point Source Catalog. Those
objects with 2MASSW designations are part of the working point
source database \citep{cut00}.}  This object was an L dwarf
candidate discovered serendipitously to be nearby on the sky to HD
89744Ab, an F7 IV-V + giant planet system.  Motivated by this
discovery, we will add all stars with known planets to the list of
candidate primaries for the dedicated search.

2MASS survey observations and evidence of common proper motion are
discussed in \S 2.  Red-optical (0.65--1.00 $\micron$) and
near-infrared (NIR) spectroscopic follow-up observations are
presented in \S 3, along with a grid of known late-M and L-dwarf
NIR spectra for classification in $z'JHK$.\footnote{\textit{\'{z}}
is a SDSS photometric band with $\lambda_c = 0.91\micron$ and
$\Delta\lambda = 0.12\micron$.  The long wave cut-off is
determined by CCD sensitivity \citep{fuk96}.} Age estimates for
the candidates are assigned in \S 4 based upon known physical
characteristics of their primaries. We use these age assignments
to determine masses for the candidates using theoretical
evolutionary curves. A discussion follows in \S 5.

\section{2MASS Observations}

\subsection{Candidate Selection}\label{cand}
The 2MASS Working Database was used to search for color-selected
low-mass candidates in close proximity to nearby stars.  The
color-selection of late-M and L-dwarfs from NIR surveys is now
well established (\citet{kir99}; hereafter K99 and \citet{kir00};
hereafter K00).  Because of their low effective temperatures, they
possess extreme visual-IR colors, e.g. $R-K_s \geq 5.5$. With the
exception of the nearest examples, late-M and L-dwarf infrared
candidates will have no optical counterparts on visual sky-survey
plates. $<\!J-K_s\!\!>$ for these objects reddens monotonically
(allowing for cosmic scatter) from $\sim1.1$ for M8 V to a maximum
of $\sim2.1$ for late L-dwarfs using classifications from the
red-optical (K99, K00).

\begin{figure}
\epsscale{0.7}
\plotone{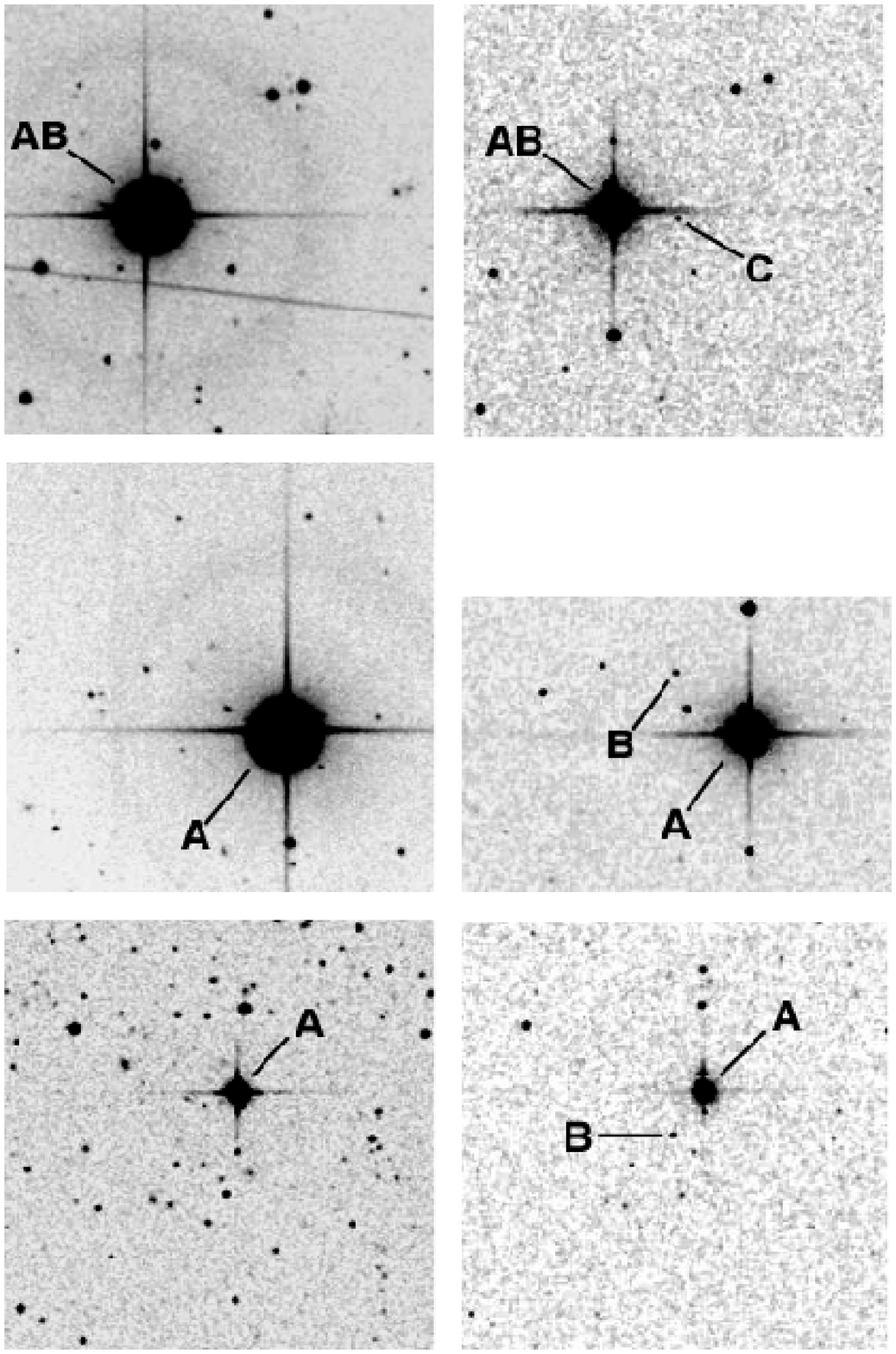}
\caption{Finder charts for all three systems.
Images are 5 arcminutes on a side with north up and east to the
left. Left panels show DSS red-band images for Gl 337 (epoch
1987.00), HD 89744 (epoch 1990.22), and Gl 618.1 (epoch 1985.44).
The primary stars are marked but the new L dwarf companions are
invisible here. Right panels show the 2MASS K$_s$-band images for
Gl 337 (epoch 2000.32), HD 89744 (epoch 1998.26), and Gl 618.1
(epoch 2000.41) on which both primary stars and L dwarf companions
are labeled. The latent images appearing $\sim$82\arcsec\ due
south of Gl 337AB and due north of HD 89744A and Gl 618.1A, as
well as the triangular-shaped reflection $\sim$11\arcsec\ due
north of Gl 618.1A, are well characterized bright-star artifacts
in 2MASS and should not be confused with real
sources.\label{finders}}
\end{figure}

The three wide separation companion candidates followed-up in this
paper met the color-selection criteria above, were in close
proximity on the sky to known nearby stars, and were bright enough
$(K_s \la 14)$ for NIR spectral follow-up with the Cornell
Massachusetts Slit Spectrograph (CorMASS; \citet{wi01a}) on the
Palomar 60-inch telescope.  The objects' coordinates, photometric
magnitudes and $J-K_s$ colors are listed in Table \ref{objects}.
Finding charts are presented in Figure \ref{finders}.

\placetable{objects} \placefigure{finders}

The companion candidate 2MASSW J0912145+145940 (hereafter
2M0912+14) has $J-K_s=1.68$ and $K_s=14.02$. It has an angular
separation of $43\arcsec$ (881 AU) from the binary Gl 337AB (Fin
347, HIP 45170),\footnote{The Washington Double Star Catalog
identifies Gl 337AB as Fin 347Aa, and lists it as the primary in
an astrometric multiple system.  The other components are BUP 125
(Aa-B), STT 569 (Aa-C), and SLE 478 (Aa-D). Comparisons of POSS-I
and POSS-II images show the three other components to be
background objects without common proper motion to Gl 337AB. Aa-B
and Aa-D appear to be the same object, simply observed at epochs
1907 and 1984, respectively.  At epoch 1984 Aa-C was located
$80.1\arcsec$ from Gl 337AB at position angle ($\phi=118\degr$)
and Aa-D was located $208.8\arcsec$ away with $\phi=208.8\degr$
\citep{mas01}.} which has an Hipparcos measured distance of
$20.5\pm0.4 \; \rm{pc}$ \citep{per97}. At this distance the
companion would have $M_{K_s}=12.47$.  Using the absolute
magnitude --- spectral class relation of K00,
\begin{equation} \label{subclass}
M_{K_s}=10.450+0.127(\rm{subclass})+0.023(\rm{subclass})^2
 \end{equation}
where subclass is $-1$ for M9 V, $-0.5$ for M9.5 V, 0 for L0 V,
0.5 for L0.5 V, etc., we infer a candidate companion spectral type
of L7 V.  K00 found $<\!J-K_s\!\!>\;\:=1.94$ with
$\sigma_{<J-K_s>}=0.37$ for the L7 subclass, so the color of
2M0912+14 is about 1 $\sigma$ from the mean.

2MASSW J1620261-041631 (hereafter 2M1620-04) has $J-K_s=1.72$ and
$K_s=13.59$. It has an angular separation of $35.9\arcsec$ (1090
AU) from the M0 V Gl 618.1 (G 17-11, HIP 80053), which has an
Hipparcos measured distance of $30.3\pm2.4 \; \rm{pc}$
\citep{per97}. At this distance the companion would have
$M_{K_s}=11.53$.  Using Equation (\ref{subclass}) we infer L4 V.
K00 found $<\!J-K_s\!\!>\;\:=1.87$ with $\sigma_{<J-K_s>}=0.16$
for the L4 subclass, so the color of 2M1620-04 is also about 1
$\sigma$ from the mean.

2MASSI J1022148+411426 (hereafter 2M1022+41) has $J-K_s=1.27$ and
$K_s=13.62$. It has an angular separation of $63.1\arcsec$ (2460
AU or 0.012 pc) from the F7 IV-V star HD 89744 (HR 4067, HIP
50786), which has an Hipparcos measured distance of $39.0\pm1.1 \;
\rm{pc}$ \citep{per97}. At this distance the companion would have
$M_{K_s}=10.7$.  Using Equation (\ref{subclass}) we infer L1 V.
K00 found $<\!J-K_s\!\!>\;\:=1.43$ with $\sigma_{<J-K_s>}=0.21$
for the L1 subclass, so the color of 2M1620-04 is within 1
$\sigma$ of the mean.

\subsection{Companionship}
The proper motions of the primaries Gl 337AB and Gl 618.1 are
$0.58\arcsec \; \rm{yr^{-1}}$ and $0.42\arcsec \; \rm{yr^{-1}}$
\citep{per97}. The 2MASS survey astrometric uncertainty is $\la
0.15\arcsec$ (1 $\sigma$; \citet{cut00}) so these motions allow
confirmation that the candidates are indeed proper motion
companions by comparing two epoch 2MASS survey images separated by
at least a few months.

The field containing 2M0912+14 was scanned twice with 2MASS, once
on 1997 November 18 and again on 2000 April 26. The scans showed
the position of 2M0912+14 to move $0.64 \arcsec \; \rm{yr^{-1}}$
towards position angle $\theta=299 \deg$ relative to other stars
in the field.  This relative motion compares well with the Gl
337AB Hipparcos absolute proper motion of $0.58\arcsec \;
\rm{yr^{-1}}$ with $\theta=295.1\deg$, thus confirming common
proper motion (CPM) companionship. Hereafter we refer to 2M0912+14
as Gl 337C.

The field containing 2M1620-04 was scanned twice with 2MASS, once
on 1999 July 29 and again on 2000 May 30.  The scans showed the
position of 2M1620-04 to move $0.44 \arcsec \; \rm{yr^{-1}}$
towards position angle $\theta=269 \deg$ relative to other stars
in the field.  This relative motion compares well with the Gl
618.1 Hipparcos absolute proper motion of $0.42\arcsec \;
\rm{yr^{-1}}$ with $\theta=273.1\deg$, thus confirming CPM
companionship. Hereafter we refer to the primary as Gl 618.1A and
2M1620-04 as Gl 618.1B.

The proper motion of HD 89744 is only $0.18\arcsec \;
\rm{yr^{-1}}$ \citep{per97}, so a longer baseline is needed to
confirm CPM companionship.  This field was scanned once with 2MASS
on 1998 April 05. The Palomar Observatory Sky Survey II (POSS-II)
also weakly detected 2M1022+41 in the photographic NIR passband
(IVN plate $+$ RG9 filter
--- $\lambda_{eff}\sim8500$ \AA; \citet{rei91}) on 1998 December
30. Unfortunately, the nine month baseline between the 2MASS and
POSS-II images is not sufficient to establish common proper
motion.

We can, however, assess the probability that 2M1022+41 is an
unassociated low-mass object based upon the likelihood that an
L-dwarf such as this object would be found in the same line of
sight as HD 89744. The surface density on the sky of 2MASS
L-dwarfs to $K_S=14.7$ is 1 per 20 square degrees (K99). There are
$\sim 6900$ stars within 50 pc listed in the Hipparcos Catalogue;
the SIMBAD database contains $\sim 7300$.  Using the larger of the
two estimates, less than one 2MASS L-dwarf will be randomly found
within $65\arcsec$ of \it{any} \rm of the $\sim 7300$ cataloged
stars. In addition, the correlation between the estimated spectral
type of 2M1022+41 from $K_S$ and the distance of the suspected
primary, L1 V (\S \ref{cand} above), and its actual spectral type
of L0 V from red-optical spectroscopy (\S \ref{os} below) strongly
supports companionship.  Based on these arguments we consider it
safe to assume companionship and hereafter refer to the primary as
HD 89744A and 2M1022+41 as HD 89744B. \citet{kor00} have
discovered a planet circling the F star, and we shall refer to
this planet as HD 89744Ab.

\section{Spectroscopic Follow-up}

\subsection{Optical Spectroscopy}

\subsubsection{Observations}\label{os}
Red-optical spectra of the candidates were obtained with the Low
Resolution Imaging Spectrograph (LRIS; \citet{oke95}) at the 10 m
W. M. Keck Observatory (Keck I).  A 400 groove mm$^{-1}$ grating
blazed at 8500 \AA \hspace{0.03cm} with a $1\arcsec$ slit and a
Tektronics $2048 \times 2048$ CCD were used to yield $1.9$ \AA
\hspace{0.03cm} pixel$^{-1}$ ($9$ \AA \hspace{0.03cm} resolution)
spectra spanning the wavelength range 6300 -- $10100$ \AA. An
OG570 blocking filter was utilized to exclude second order.
Further details regarding data acquisition and reduction are given
in K99 and K00. Table \ref{observations} summarizes observations.

\placetable{observations}

\begin{figure}
\plotone{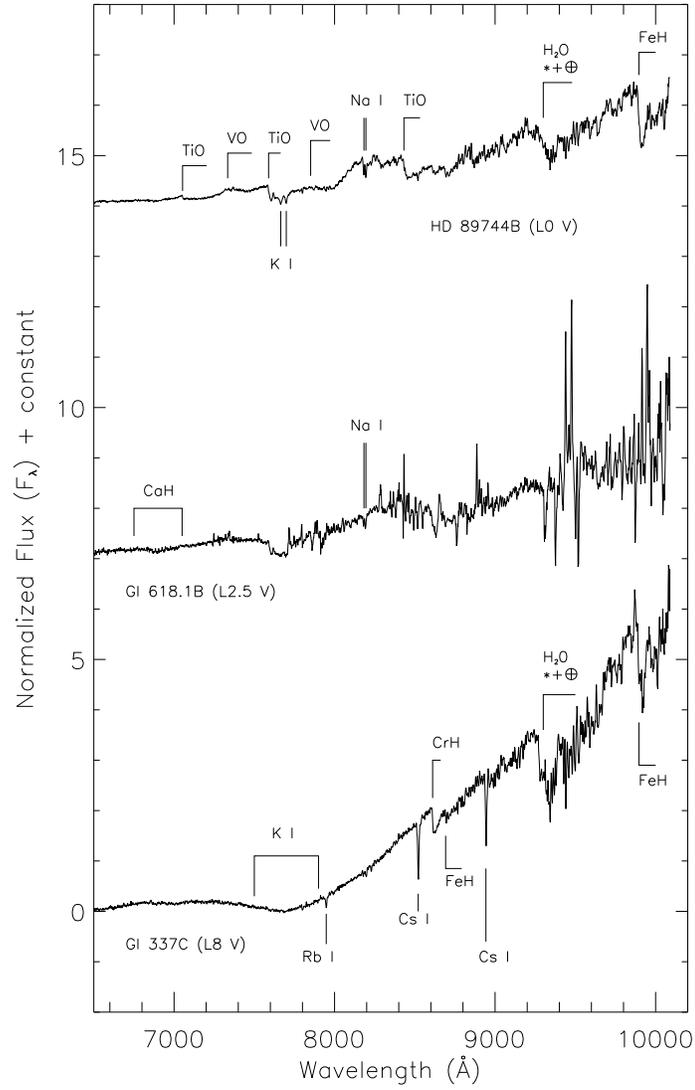}
\caption{Red-optical spectra of the three
L-dwarf companions from Keck I.  The flux scale is in units of
$F_\lambda$ normalized to one at $8250$ \AA.  Integral offsets
have been added to the flux scale to separate the scale
vertically. Hallmark features of the L-dwarf class are identified.
The Gl 618.1B is very noisy due to observations through cirrus and
between episodes of fog.\label{optical spectra}}
\end{figure}

Figure \ref{optical spectra} displays the reduced optical spectra.
The spectra have not been corrected for telluric absorption.  They
confirm that the objects are L-dwarfs. Using the L-dwarf
classification scheme of K99 we assign the following spectral
types: Gl 337C (L8 V), Gl 618.1B (L2.5 V), and HD 89744B (L0 V).
These spectral types correspond to effective temperatures of L8 V
($1300-1600$ K), L2.5 V ($1800-1950$), and L0 V ($2000-2200$),
where the lower and upper limits are from the classifications
schemes of K00 and \citet{bas00}, respectively.

\placefigure{optical spectra}

\subsubsection{Lithium}\label{lithium}
The \ion{Li}{1} ($6708$ \AA) feature is an important indirect
gauge of mass \citep{reb92}. When seen in absorption in all but
the youngest fully convective very low-mass objects, this feature
indicates insufficient mass ($M \la 60 M_{jup}$, i.e. sub-stellar)
to produce central temperatures high enough ($T_{cen}
> 2 \times 10^6 K$) to fully fuse the primordial lithium through
$(p,\alpha)$ reactions (see e.g. \citet{nel93}).  Lithium is not
detected in any of the objects to the following equivalent width
(EW) upper limits: $< 1$ \AA \hspace{0.03cm} for Gl 337C and Gl
618.1B, and $< 0.5$ \AA \hspace{0.03cm} for HD 89744B. This
indicates that Gl 618.1B and HD 89744B have $M \ga 60 M_{jup}$, an
important mass constraint on the secondaries that we will utilize
in \S \ref{primasses}.

Gl 337C is too cool to use the lithium test.  As temperatures fall
in the atmospheres of L-dwarfs, lithium is expected to give way to
LiCl as the dominant lithium bearing gas for $T_{eff} \la
1500-1550K$, thereby depleting existing atomic lithium
\citep{lod99}.  Since our assigned $T_{eff}$ upper limit for Gl
337C (1600 K) is only slightly warmer than this transition
temperature, we consider the use of this test misleading for this
object.  K00 found diminishing Li absorption line strengths for L7
and L8 dwarfs when present, and interpreted this as evidence of
increasing lithium bearing molecular formation.

LiCl spectral band heads occur in the mid-infrared at
$14.51\micron$ ($\nu=0-1$), $14.71\micron$ ($\nu=1-2$), and
$14.92\micron$ ($\nu=2-3$) \citep{kle60}.  Unfortunately, these
wavelengths are not accessible from the ground due to atmospheric
absorption just beyond the $N$ passband, but they will be
observable with the Infrared Spectrograph (IRS) on the upcoming
Space Infrared Telescope Facility (SIRTF).

\subsubsection{$\rm{H}\alpha$}
$\rm{H}\alpha$ seen in emission is an important measure of
chromospheric activity (and youth) in early-mid M-dwarfs, just as
\ion{Ca}{2} H and K lines are used for FGK stars.  $\rm{H}\alpha$
emission is not detected with upper limits $<1$ \AA
\hspace{0.03cm} for Gl 337C and $< 0.5$ \AA \hspace{0.03cm} for HD
89744B.  Gl 618.1B is a possible detection with $\rm{EW}_{\rm{Li}}
\leq 1$ \AA. These results are compatible with the L-dwarf
activity statistics from the K99 and K00 samples, in which
$\rm{H}\alpha$ emission declined from 60\% (80\% if marginal
detections are included) for L0 V to 8\% (25\% including two
marginal detections) for L8 V \citep{giz00}. Statistically we
might have expected emission in HD 89744B, but it is not
surprising that Gl 618.1B and Gl 337C are not active.

\subsection{Near-Infrared Spectroscopy}
The wavelength of peak energy emission shifts redward into the NIR
with decreasing effective temperature, so L-dwarfs become very
faint in the red-optical. As noted in K99, a large aperture
telescope is needed to use red-optical spectral typing schemes
effectively. Because a more practical means of assigning spectral
types to color-selected late-M and L-dwarfs was sought, the
Cornell Massachusetts Slit Spectrograph (CorMASS; \citet{wi01a})
was built for use on the Palomar 60-inch telescope.  Over 100 new
field late-M and L-dwarf candidates have been observed with
CorMASS as part of a magnitude-limited spectroscopic survey to
improve the statistics of the luminosity function across the
stellar---sub-stellar boundary in the solar neighborhood.  The
instrument has also been used for the spectral confirmation of a
bright T-dwarf \citep{bur00} and now for the follow-up of some
candidate wide separation companions in \citet{gi01b} and this
paper.

CorMASS is a prism cross-dispersed low-resolution ($R\sim300$)
near-infrared (NIR) spectrograph. Its 2-dimensional spectral
format provides simultaneous coverage from $\lambda\sim 0.75
\micron$ to $\lambda \sim 2.5 \micron$ (\textit{z'JHK} bands).  A
40 lines mm$^{-1}$ grating, blazed at $4.8\micron$, is used with a
fixed $2 \arcsec$ slit to image the raw spectra across 6 orders on
a NICMOS\,3 detector.

\begin{figure}
\plotone{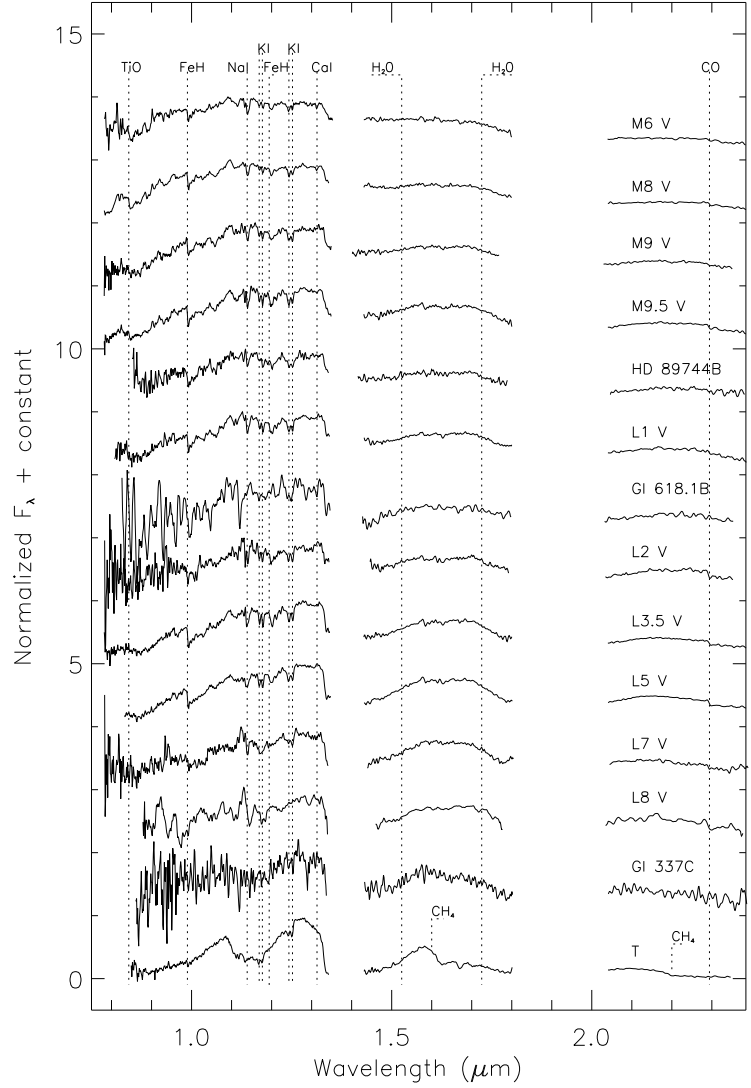}
\caption{A grid of NIR spectra of known
low-mass objects with spectral types M6 V -- T. Important spectral
features are indicated. Data between bands was not usable owing to
poor atmospheric subtraction. Spectra have been normalized at the
J-band peak and integral flux offsets have been added.  The L2 V,
L7 V, and L8 V have poor signal-to-noise shortward of $\sim 1.1
\micron$. Spectra of Gl 337C, Gl 618.1B and HD 89744B appear in
the grid at positions bracketed by grid objects with similar
features.  The spectrum of Gl 618.1B has been smoothed with a
boxcar average of 5 pixels.\label{NIR grid}}
\end{figure}

Figure \ref{NIR grid} presents a grid of CorMASS NIR spectra for
known low-mass objects spanning spectral classes M6--T.  An
observing log for the grid is presented in Table \ref{grid log}.
The M-dwarf spectral types are on the \citet{kir91} system, and
L-dwarf classifications are on the K99 system.  The data reduction
technique follows that described in \citet{wi01a}.  Our grid
qualitatively agrees with similar NIR grids of L-dwarfs already
published \citep{re01b,tes01}.  As seen in Figure \ref{NIR grid},
the NIR spectra are dominated by molecular absorption features
that smoothly change with spectral type through the end of the
L-class. The $\rm{H_2O}$ absorption features bracketing H-band
change most obviously.  In particular, the slope of the blue side
of the H-band spectra appears to increase monotonically with later
spectral type as $\rm{H_2O}$ absorption increases in the L-dwarf
atmospheres. This monotonic relationship has been quantitatively
confirmed for a grid of L-dwarfs by \citet{re01b} and
\citet{tes01} and for a grid of M-dwarfs by \citet{jon94}.
\citet{del99} also use this feature to help classify DENIS
candidate low-mass objects.

\placefigure{NIR grid} \placetable{grid log}

Other spectral changes in the grid include varying strengths of
the FeH features at $0.99 \micron$ and 1.19-1.24 $\micron$, as
well as changes in the pair of \ion{K}{1} doublets in J-band.  The
evolution of the J-band features were investigated by
\citet{mcl00} using higher resolution spectra of L-dwarfs. K-band
evolution seen in Figure \ref{NIR grid} includes increasing
$\rm{H_{2}O}$ absorption on the blue end, increasing curvature
longward of $\sim 2.1 \micron$ due to collision induced absorption
of $\rm{H}_2$, and a deepening of the CO bandhead feature at $\sim
2.3 \micron$ (see e.g. \citet{tok99}). The slow spectral evolution
for late-M and L-dwarfs is in marked contrast to the rapid change
in features observed from L8 V to T dwarfs as the primary
carbon-bearing molecule changes from CO to $\rm{CH_4}$ .

CorMASS observations of the companions are listed in Table
\ref{observations}.  The spectra have been inserted into Figure
\ref{NIR grid} at positions bracketed by known objects with
similar spectra based upon comparisons by eye.  The NIR spectra of
Gl 337C supports a spectral type assignment of L8 V. The depth of
the J-band $\rm{H_2O}$ ($\sim 1.15\micron$) absorption feature
indicates a spectral type later than L7 V.  The spectra of Gl
618.1B (smoothed with a boxcar average of 5 pixels) and HD 89744B
are both similar to those of the late-M and early L-dwarfs.
Because the spectral features evolve more slowly in this regime we
cannot confirm a specific spectral type by eye.  The spectra
support a range of spectral types: M9 V $> \rm{Sp T}_{\it{Gl
618.1B}} >$ L3.5 V and M9 V $> \rm{Sp T}_{\it{HD 89744B}} >$ L2 V.

At the resolution of CorMASS (R $\sim 300$) quantitative
comparison of spectral type indices (i.e. ratios of average flux
in specified regions, or equivalent widths, etc.) will be
necessary to improve the accuracy of spectral subclass assignments
based upon NIR spectra.  In a subsequent paper \citep{wi01b} we
will define quantitative spectral indices for assignment of
spectral subclass for use with the CorMASS spectroscopic survey of
field low-mass object candidates.

\section{Age Estimates}\label{ages}
The assignment of ages to field stars is by its very nature
imprecise due primarily to few reliable techniques, cosmic
scatter, and especially slow evolution for stars with late
spectral types. Hopefully the analysis of systems with late-M and
L-dwarf wide separation companions will motivate increased study
of `age dating' techniques. This necessarily lengthy section
gathers available evidence and applies relevant techniques to
derive unfortunately imprecise age estimates for the primary
systems of the L-dwarf companions.

Published age related properties for the three primary systems are
summarized in Table \ref{primproperties}.  A decrease in activity
(coronal and chromospheric) with age in stars with convective
envelopes is generally attributed to a reduction in the classical
dynamo effect ($\alpha$-$\Omega$) as stars spin down due to
angular momentum losses in stellar winds (see e.g. \citet{bal85})

\placetable{primproperties}

Surface lithium abundance from high resolution spectroscopy can
serve as a crude age indicator for main sequence stars with
convective envelopes.  Primordial surface lithium in these stars
is destroyed through mixing to warmer interior temperatures due to
the combined action of convection in the outer envelope and more
complex mechanisms not yet fully understood \citep{pin97}. Only
one of the three primaries in this paper have published lithium
abundance measurements, namely HD 89744A. Unfortunately, the
measurement's usefulness as an age diagnostic is diminished
because, as an F7 IV-V, it falls very close to the mid-F `lithium
dip' region of the main sequence in which dramatic surface lithium
destruction occurs where standard evolutionary models using
convective mixing predict there should be no destruction (see,
e.g. \citet{ba95}).

Space motions are frequently used as indicators of age for samples
of stars. But their use in assigning ages to individual stars must
be done with great caution due to significant dispersion within
any kinematic class. Nonetheless, it is unlikely for an old star
to have small velocity, but a young star can have a high velocity.
We take the division between \it{young} \rm{and} \it{old} \rm to
be $\sim 1.5-2$ Gyr, the division between the young disk and old
disk used by \citet{eg89b}.  We will refrain from placing
significant weight on kinematic age when possible.

In the case of HD 89744A there are several explicit age
determinations already published based upon high resolution
metallicity measurements and fits to stellar isochrones. These
ages will be compared with our estimates derived from physical
properties.

\subsection{Gl 337 System}
Gl 337AB is a double-lined spectroscopic (SB2) and visual binary
located $20.5\pm0.4$ pc away (Hipparcos, \citet{per97}).
Astrometric, speckle and visual observations reveal an orbital
semi-major axis of $\sim0.116\arcsec$ ($\sim2.4$ AU) and a period
of 2.7 years (see e.g. \citet{pou00,mas96}). It is one of the
shortest-period visual pairs observed.  Published measurements are
summarized by \citet{mas96}. They adopted a spectral type of G8 V
for both members based on $ubvy$ photometry. More recent lunar
occultation \citep{ric00} and adaptive optics observations
\citep{bar00} indicated an early K spectral type for the secondary
based on $\sim0.1$ brighter magnitudes than the primary at $K$ and
$r$ bands.

\subsubsection{Age from Coronal Activity}\label{alphasource}
Gl 337AB is detected as a ROSAT X-ray source with $L_X = 9.3
\times 10^{27} \; erg \; s^{-1}$ \citep{hun99}. The quantity
$R_X=\log{{L}_X/{L}_{bol}}$ is often used to describe X-ray
emission in a distance and stellar radius independent logarithmic
ratio. \citet{gai98} has derived a relation between $R_X$ and age
for solar type stars using published empirical relations for
rotational period time dependence, X-ray luminosity/rotation
correlations, and the luminosity evolution of the constant-mass
Sun:
\begin{equation} \label{xray}
R_{X}=-6.38-2.64\alpha\log{(t/4.6)}+\log{[1+0.4(1-t/4.6)]}
\end{equation}
The coefficient $\alpha$, from the rotation period time dependence
relation, has published values of $1/2$ \citep{sku72} and $1/e$
\citep{wal91}. The age $t$ is in Gyr. Following \citet{kir01} we
have adjusted the Equation to give $R_{X}=-6.38$ \citep{mag87} for
the present solar age of 4.6 Gyr.

Because the individual stars of Gl 337AB are not resolved by
ROSAT, and they are close in spectral type, we attribute \it{ad
hoc} \rm half of the observed X-ray luminosity to each star. Using
$L_{X, Gl 337A} = 4.65 \times 10^{27} \; erg \; s^{-1}$ and
$L_{bol, Gl 337A}=0.64\pm0.06L\sun$ \citep{mas96}, we calculate
$R_{X}=-5.72$. Equation (\ref{xray}) can then be used to estimate
an age of 1.25 Gyr ($\alpha=1/2$) and 1.7 Gyr ($\alpha=1/e$) for
Gl 337AB.

This estimate implicitly assumes that the total X-ray luminosity
measured by ROSAT is merely the sum of individual X-ray emission
from \it{non-interacting} \rm members of a binary system. This
assumption is valid. \citet{duq92} reviewed physical modes of
interaction in binaries with solar-mass primaries. Since both
members of Gl 337AB are on the main sequence, interaction modes
typical of evolved stars such as stellar wind accretion and
Roche-lobe overflow can be rejected. Tidal interactions can
circularize and synchronize binary orbits for systems on the order
of $P\sim10$ days, i.e. close binaries, and this interaction is
thought to cause the $P$ cut-off below which all close binaries
have circular and presumably synchronized orbits (see e.g. DM91
Figure 5). Tidal interaction in close binaries nullifies the use
of X-ray emission as an indicator of age since the tidal effects
disrupt the spin-down, and thus rotation-activity correlation, of
the individual stars.

But timescales for orbital synchronization $(t_{sync})$ and
circularization $(t_{circ})$ for stars with convective envelopes
are extremely steep functions of binary fractional separation
$(a/R)$: $t_{sync} \propto (a/R)^6$ and $t_{circ} \propto (a/R)^8$
\citep{zah77}. \citet{zah77} estimated $t_{sync} \sim
10^4((1+q)/2q))^2P^4$ years, where q is the mass fraction and P is
the orbital period in days.  Using $q \sim 0.96$ and $P \sim 985$
days for Gl 337AB gives $t_{sync} \sim 10^{16}$ years.  Gl 337AB
is therefore not a tidally interacting binary.

\subsubsection{Kinematics}
The space motion of Gl 337AB is directed opposite the Galactic
center: $\rm{U}=-74$ km $\rm{s^{-1}}$, $\rm{V}=-2$ km
$\rm{s^{-1}}$, and $\rm{W}=+3$ km $\rm{s^{-1}}$ \citep{egg98}.
(This is with respect to the sun and with U positive towards the
galactic center.)  This motion makes it a member of the kinematic
old disk population with an age of $\sim2$ to 10 or 12 Gyr based
upon its position outside of the young disk `Eggen box' in the UV
plane \citep{eg89b}.  The `Eggen box' is roughly defined by $-50$
km $\rm{s^{-1}}$ $< \rm{U} < +20$ km $\rm{s^{-1}}$ and $-30$ km
$\rm{s^{-1}} < \rm{V} < 0$ km $\rm{s^{-1}}$.

\subsubsection{Age from Isochrone Comparisons}
Since the Gl 337AB system is both an SB2 and visual binary, mass
and magnitude measurements for both components have been
published. If we assume that the stars of the binary are coeval,
plotting the observed magnitudes against model evolutionary
cooling curves \it{individually} \rm for the components should
indicate the same age.

We assumed the mass estimates of $M_A=0.89\pm0.029M_{\sun}$ and
$M_B=0.85\pm0.026M_{\sun}$ \citep{pou00} and absolute magnitudes
of $M_{K_{FIRPO}}=3.94$ and $M_{K_{FIRPO}}=4.04$, inferred from
lunar occultation measurements \citep{ric00} and Hipparcos
distances, for the components. The measured absolute magnitudes
were transformed to the appropriate $K$ passbands for comparison
with the theoretical zero-metallicity evolutionary models ($M_K$
vs. age) of \citet{bar98} and \citet{gir00}.

Surprisingly, we found disparate ages of roughly 1 Gyr and 3 Gyr
for components A and B, respectively, when compared to both
models.  The most obvious discrepancy was the factor of 2
difference between the observed ($\Delta M_{K_{CIT}}=0.10$) and
predicted ($\Delta M_{K_{CIT}} \sim0.2$) differential magnitudes
for mass differences of $\sim 0.05M_\sun$. Assuming larger
published mass differentials for the components increases the
discrepancy. There are no other published $K$ band measurements.

It should be noted that the masses of the components fall within
an area of extreme sensitivity to model mixing parameter,
metallicity, and initial He abundance, as well as other model
assumptions (I. Baraffe 2001, private communication).  These
complexities likely contribute to the age disparity, but it is
unlikely that they fully account for the discrepancy given in
particular the \citet{bar98} model's success in other
empirical--theoretical comparisons (see e.g. \citet{del00}). While
we cannot use these isochrone comparisons to constrain an age for
the Gl 337 system, the discrepancy highlighted by this comparison
is worthy of further study both from the standpoint of theoretical
modelling and via observation.

\subsubsection{Adopted Age}
We adopt an age span of 0.6 to 3.4 Gyr by simply assuming a factor
of 2 uncertainty in the 1.25 -- 1.7 Gyr estimate derived from
X-ray luminosity.  Our adopted age is consistent with the
kinematic age if the Gl 337 system is a younger member of the old
disk.

\subsection{Gl 618.1 System}
Gl 618.1A is an M0 V star located $30.3\pm2.4$ pc away based on
Hipparcos measurements.  This particular object has not been
extensively studied.  It is worth noting that field M-dwarfs are
notoriously difficult to age date for various reasons, including:
1) slow evolution on the HR diagram in a Hubble time, 2) fast
depletion of photospheric lithium since full convection appears as
early as M3 V, 3) generally slow rotation, 4) poorly correlated
rotation and activity strength \citep{rh00}, and 5) intrinsic
faintness.

\subsubsection{Activity}
Gl 618.1A does not display evidence of chromospheric or coronal
activity: ROSAT did not detect X-ray emission from the vicinity of
this object, and $\rm{H}\alpha$ was observed in absorption on two
occasions separated by a decade \citep{sod85,rei95}.  It would
actually be unusual if Gl 618.1A displayed evidence of activity
since only 8\% of 98 M0 V dwarfs observed by \citet{haw96} were
magnetically active (dMe).\footnote{A dMe is defined as a star
with $\rm{EW}_{\rm{H} \alpha}>1$ \AA.}  \citet{haw96} found that
in a sample of nearly 600 nearby M-dwarfs, the dMe stars ($\sim
20\%$ of the sample) came from a kinematically younger population
than the remaining dM stars in the survey.

\subsubsection{Previous Age Analysis}
\citet{leg92} classified over 200 low-mass stars into five
populations using kinematic and photometric techniques: young
disk, young/old disk, old disk, old disk/halo, and halo.  The
objects in the sample were first placed into populations based
upon kinematics. Gl 618.1A's space motion of $\rm{U}=+122.3$ km
$\rm{s^{-1}}$, $\rm{V}=-62.2$ km $\rm{s^{-1}}$, and $\rm{W}=-12.8$
km $\rm{s^{-1}}$ met the criteria for membership in the old
disk/halo based upon its eccentricity of $\sim 0.5$ in the U-V
plane.  \citet{leg92} also utilized infrared color-color diagrams
to estimate mean metallicity ($\rm{[m/H]}$).  The kinematic
populations occupied separate positions on the diagrams, albeit
with scatter.  Despite its space motion, Gl 618.1A was found to be
a member of the young disk on this diagram and was assigned
$\rm{[m/H]} \sim 0$. The sample photometric uncertainties gave a
class membership uncertainty of $\pm 1$. Lastly, \citet{leg92}
plotted the objects in color-absolute magnitude diagrams and again
found the populations could be predicted based upon location. This
diagram also supported Gl 618.1A's membership in the young disk.

\subsubsection{Adopted Age}\label{gl 618.1b age}
We are confronted with conflicting and low confidence age
diagnostics for this object: large total space motion (138 km
$\rm{s^{-1}}$) that is unlikely to belong to a young disk ($t<1.5$
Gyr), and photometric evidence of youth \citep{leg92}.  We refrain
from adopting a lower limit, and simply allow the lower limit to
be set by our non-detection of lithium in the spectrum of the
L-dwarf companion (\S \ref{lithium}).  This imposes a lower limit
of $\sim 0.5$ Gyr (\S \ref{primasses}). We adopt an upper limit
age of 12 Gyr based on the recently determined solar neighborhood
age of $11.2\pm0.75$ Gyr \citep{bin00}.

\subsection{HD 89744 System}
HD 89744A is an F7 star at $39.0\pm1.1$ pc with ${\rm M}_{\rm V} =
2.79\pm0.06$ based upon Hipparcos measurements \citep{per97}.
\citet{kor00} summarized this star's properties when they reported
the discovery of a massive planetary companion to HD 89744A. The
authors adopted a mass of $M=1.4\pm0.09M_{\sun}$ for HD 89744A
based upon the average of independent results from \citet{all99}
($M=1.34\pm0.09M_{\sun}$) and \citet{ng98}
($M=1.47\pm0.01M_{\sun}$).

The massive planet HD 89744Ab ($m \sin{i} = 7.2 M_{jup}$) orbits
the primary every 256 days with a highly eccentric orbit
($\rm{e}=0.7$). The rms residual uncertainty to the orbital fit of
the massive planet was 20.5 m $\rm{s}^{-1}$ \citep{kor00}. This
residual exceeded the instrumental uncertainties of 10 m
$\rm{s}^{-1}$ and led the authors to state that they could not
rule out the existence of a distant companion.  A crude estimate
of the velocity of the massive planet HD 89744Ab induced by the
wide L-dwarf companion HD 89744B is $v \sim a \times \triangle
t_{data} = (\rm{G}\it m/r^2) \times \triangle t_{data}$, where $a$
is the acceleration of the planet due to the gravitational force
of the L-dwarf, $\triangle t_{data}$ is the $\sim 3$ year time
span of \citet{kor00} radial velocity observations, $m \sim
0.08M_\sun$ is the mass of L-dwarf, and $r=2500$ AU. This gives $v
\sim 0.007$ m $\rm{s}^{-1}$, thus ruling out the possibility that
HD 89744B is the cause of the orbital fit residuals.  At 2500 AU
separation, a companion must be $m \sim 100M_\sun$ to induce a
residual radial velocity equal to the instrumental uncertainty of
the \citet{kor00} observations. Of course this does not rule out
the possibility of another as yet unseen companion at a smaller
orbital separation.

\subsubsection{Ages from Isochrone Fitting}
The age of HD 89744A has been estimated from fits to evolutionary
isochrones on three occasions. \citet{edv93} calculated an age of
2.09 Gyr using $\rm{[ Fe/H}]=0.18$.  The object fell within the
`hook-region' of the isochrones used by \citet{edv93}, so the
authors noted that the age could be underestimated by 0.15 dex,
corresponding to an older age of 2.95 Gyr. \citet{ng98} also
calculated an age with this metallicity but used more recent
stellar isochrones and Hipparcos distances. They gave an age of
2.04 Gyr. And \citet{gon01} derived an age of 1.8 Gyr from
isochrone fits. All three ages are in close agreement.

\subsubsection{Luminosity Class}
\citet{saa99} revised this object's luminosity class from
main-sequence to subgiant using the improved ${\rm M}_{\rm V}$ and
hence more accurate location in color magnitude diagrams afforded
by Hipparcos results.  This change is supported by the finding of
\citet{edv93} that HD 89744A was within the `hook-region'.
\citet{bar96} assigned a spectral type of F7 IV-V using the
Vilnius Photometric System.  \citet{ng98}, using updated
isochrones and Hipparcos distances as well, still found the object
to be in luminosity class V. Due to the uncertainty in luminosity
class we will refer to the object as an F7 IV-V. Based on the
object's proximity to the main sequence turn-off, its age can be
estimated to be $\sim2$ Gyr, the approximate turn-off age of an
$M=1.4M_{\sun}$ star with our adopted metallicity
($[\rm{Fe/H}]=0.24$; Table \ref{primproperties}) from the
\citet{ber94} isochrones used by \citet{ng98}.

The projected rotational velocity ($v \sin{i}$) of 8 km
$\rm{s^{-1}}$ \citep{hof91} for HD 89744A supports a spectral type
of F7 IV-V. This rotation falls squarely in the 0 to 20 km
$\rm{s^{-1}}$ range found in a study of late-F subgiants by
\citet{ba90}.  More recently \citet{dem97} and \citet{leb99} found
a sharp rotational discontinuity at $\bv \sim 0.55$, or F8 IV.
The rotational velocities of subgiants earlier than F8 spanned
from a few to $\sim 180$ km $\rm{s^{-1}}$, while subgiants later
than F8 had mean rotational velocities of 6 km $\rm{s^{-1}}$ at G0
decreasing to $\sim 1$ km $\rm{s^{-1}}$ at K5. Abrupt magnetic
braking with the deepening convective envelope in the subgiant
phase is thought to be the cause of this discontinuity
\citep{leb99}. While there is significant scatter in the
rotational velocities of mid to late F-dwarfs, the $v \sin{i}$ of
HD 89744A is compatible with a subgiant evolving redward towards
the F8 IV discontinuity.

\subsubsection{Age from Activity}
\citet{piz00} reported an upper limit of $R_X < -5.7$ from ROSAT
measurements in their study of X-ray emission from early post-main
sequence stars. From Equation (\ref{xray}) this gives a minimum
coronal age of 1.2 Gyr ($\alpha=1/2$) and 1.7 Gyr ($\alpha=1/e$).
But since the X-ray measurement is an upper limit, the age
indicated by this method is likely much greater.

Another indicator of activity is $\log{(R^\prime_{HK})}$, where
$R^\prime_{HK}$ is the ratio of chromospheric emission in the
\ion{Ca}{2} H and K lines and the bolometric luminosity
\citep{noy84}.  For this indicator chromospheric emission is
measured using the Mount Wilson HK pseudo-equivalent width $S$
(see below). We can use a relation between age and $R^\prime_{HK}$
from \citet{don93} based upon observations of clusters spanning a
wide range of ages:
\begin{equation} \label{chromage}
\log(t)=10.725-1.334R_5+0.4085{R_5}^2-0.0522{R_5}^3
\end{equation}
where the age $t$ is in years and $R_5=10^5 \times R^\prime_{HK}$.
Using Equation (\ref{chromage}) with $\log{(R^\prime_{HK})}=-5.04$
\citep{sod85} gives an age of 6.4 Gyr for HD 89744A. \citet{gon01}
found $\log{(R^\prime_{HK})}=-5.12$ and derived an age of 8.4 Gyr
using Equation (\ref{chromage}) as well.

Chromospheric emission is often measured using the Mount Wilson
\ion{Ca}{2} index $S\propto[(H+K)/(V+R)]$, where $H$ and $K$ are
counts in the \ion{Ca}{2} passbands and $V$ and $R$ are counts in
violet and red continuum bands adjacent to the H-K regions
\citep{vau78}. \citet{bar88} used an empirical relationship
between the mean values of $<\!\rm{S}\!>$ and $\bv$ to derive an
age of 9.9 Gyr.

This gives age estimates from activity ranging from 6.4 to 9.9
Gyr, in marked contrast to the $\sim 2$ Gyr age from isochrone
fits. Why are the chromospheric age estimates above so much older
than ages from fits to isochrones?  Interestingly, HD 89744A is
known to have flat activity cycles (i.e. no evidence of activity
cycles such as the 11 year solar cycle).  Because of its flat
chromospheric emission, \citet{wil68} used this object as a
standard in his studies of activity variability. Over a 25 year
period the standard deviation of S ($<\sigma_S/S>$) for this
object was 1.2\% \citep{bal95}.

\citet{bal90} postulated that solar-type stars that exhibit
prolonged low levels of magnetic activity may be in Maunder
Minimum-like phases. If HD 89744A is in such a phase then measures
of activity will lead to old age estimates that are misleading.
Lastly, we note that evolutionary diagrams predict even
$1.25M_{\sun}$ stars to reach the tip of the red-giant branch in
$<5$ Gyr \citep{ibe67}; the $\sim1.4M_{\sun}$ HD 89744A would be a
white dwarf at the ages predicted by its activity.

\subsubsection{Kinematics}\label{metal}
The space motion of HD 89744A is $\rm{U}=+9.3$ km $\rm{s^{-1}}$,
$\rm{V}=-25.6$ km $\rm{s^{-1}}$, and $\rm{W}=-13.6$ km
$\rm{s^{-1}}$ \citep{edv93}.  This space motion falls within the
UV `Eggen box' corresponding to the young disk with ages $\la 1.5
- 2$ Gyr \citep{eg89b}.

\subsubsection{Adopted Age}
Based upon the well correlated published ages from isochrone fits,
as well as evidence of proximity to the main sequence turn-off, we
adopt an age of 1.5-3 Gyr for HD 89744A.  We ignore age estimates
from activity.

\section{Discussion}
\subsection{Companion Masses and Colors}\label{primasses}
\begin{figure}
\plotone{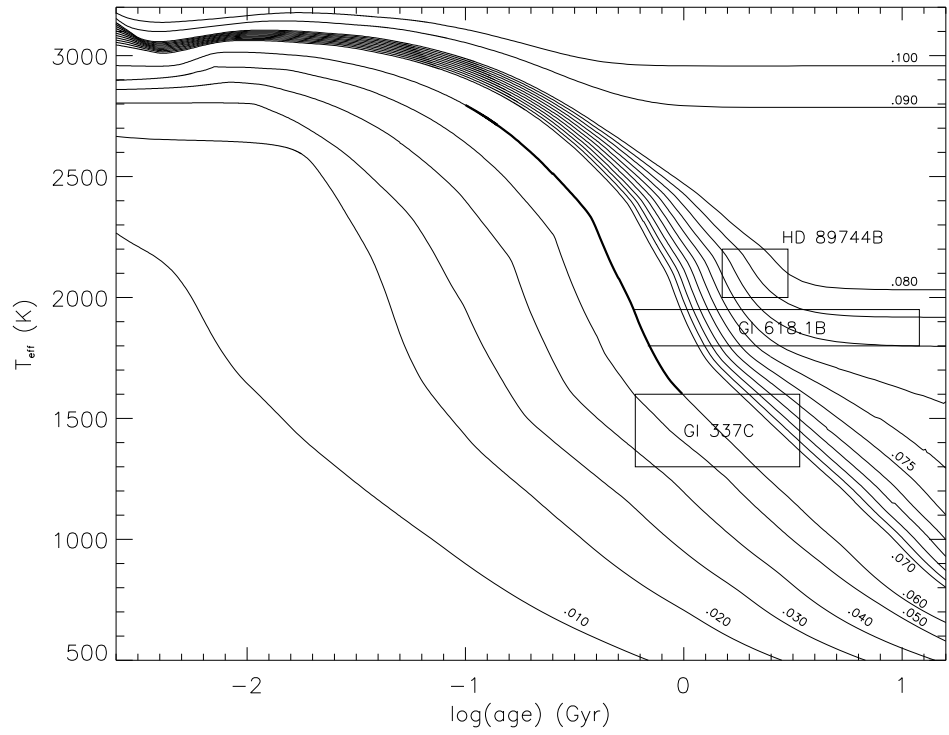}
\caption{Evolutionary models for brown dwarfs
and very low-mass stars with masses from $0.010 M_\sun$ through
$0.100 M_\sun$ from \citet{bur97}.  The approximate position of
the lithium burning cut-off, $0.060 M_\sun$, is highlighted.
Overplotted are error boxes representing the age and $T_{eff}$
estimates we derived for the three companion
L-dwarfs.\label{evolutionary curves}}
\end{figure}

With estimated ages for the L-dwarf companions from \S \ref{ages}
and effective temperature estimates from spectral types (\S
\ref{os}), the masses of the L-dwarf companions can be estimated
using evolutionary models from \citet{bur97}. In Figure
\ref{evolutionary curves} error boxes are plotted over the
evolutionary curves to represent the range of effective
temperatures and ages assigned to the objects. Based on the
lithium test results we also use the $60 M_{jup}$ curve as the
minimum mass boundary for Gl 618.1B since we were unable to
constrain a lower limit age from physical properties of this
object's primary (\S \ref{gl 618.1b age}). Masses can then be
easily determined from the figure: $40M_{jup} \leq M_{Gl 337C}
\leq 74M_{jup}$, $60M_{jup} \leq M_{Gl 618.1B} \leq 79M_{jup}$,
and $77M_{jup} \leq M_{HD 89744B} \leq 80M_{jup}$. Thus Gl 337C is
a brown dwarf, HD 89744B is a very low-mass star, and Gl 618.1B
could be either.  Properties of the L-dwarf companions are
summarized in Table \ref{secparameters}.

\placefigure{evolutionary curves} \placetable{secparameters}

\subsection{Primary Masses}
The total mass of the SB2 binary Gl 337AB is well constrained from
astrometry to be $M_\sun = 1.74\pm0.15$ \citep{pou00}.  A mass for
Gl 618.1A can be derived using its absolute $K$ magnitude and an
appropriate mass-luminosity relation (MLR) for sub-solar mass
stars. Using $K_{\rm{CIT}}=7.11$ \citep{leg92} and an Hipparcos
distance of 30.3 pc gives $M_K=4.70$.  \citet{hen93} give an
empirically determined MLR of
\begin{equation} \label{mlr}
\log{M/M_\sun}=-0.1048(M_K)+0.3217
\end{equation}
for $M_K=3.07$ to $5.94$.  Using Equation (\ref{mlr}) we calculate
$M_{\rm{Gl 618.1A}}=0.67 M_\sun$.

\citet{kor00} adopted a mass of $1.4\pm0.09 M_\sun$ for HD 89744A
from the average of published results.  These authors determined
$m\sin{i}= 7.2M_{jup}$ for the extrasolar planet HD 89744Ab. For a
large sample \citet{cha50} have shown
$<\!\sin{i}\!\!>\;\:=\pi/4=0.79$. Assuming $\sin{i}$ for the
planet's orbit is reasonably close to this expectation value, the
planet's mass will be negligible compared to that of the primary.
Hence we adopt a combined mass of $1.4M_\sun$ for the primary $+$
planet.

\subsection{System Characteristics}
Based on the primary and secondary masses derived above, upper
limit mass ratios for the Gl 337ABC, Gl 618.1AB, and HD 89744AbB
systems are 0.04, 0.12, and 0.06, respectively (Table
\ref{sysparameters}).  Of the six previously discovered systems
with wide-separation L-dwarfs, 4 of 6 have $q<0.1$. It is
interesting to ask how many of these systems meet the sample
criteria of the \citet{duq91} G-dwarf binarity study?
\citet{duq91} (hereafter DM91) specifically take as a sample F7 to
G9 stars from the Catalogue of Nearby Stars (second edition;
\citet{gli69}) with luminosity class IV-V, V, and VI, declinations
above $-15\degr$, and trigonometric parallax $>0.045\arcsec$, i.e.
stars within 22 pc. Two of the primary systems in their study, Gl
417A and Gl 584AB, have wide-separation L-dwarf companions
\citep{kir01} that were not found by the DM91 radial velocity
studies of the primaries.  One of the primaries in this paper, Gl
337AB, meets the DM91 criteria except it was not added to the
Catalogue of Nearby Stars (CNS) until the third edition in 1991
\citep{gli91} and it was assigned a combined K0 V spectral type.
(We adopted the spectral type assignment G8 V + K1 V).

\placetable{sysparameters}

Given these newly discovered systems and improvements in the
already well studied CNS, it would be useful to search for wide
binaries around a current DM91 style G-dwarf sample to improve our
understanding of the distribution of mass ratios below
$\sim0.2$.\footnote{The recent results of extrasolar planet
searches will also impact the DM91 results, although then one must
address the issue of when does a binary system change its label to
primary star $+$ planet. For instance, 51 Peg (Gl 882), with its
recently discovered extrasolar planet, is in the DM91 sample.}

Returning to the three wide-companions in this paper, we find it
interesting to compare these systems to other binaries in
($\log{\Delta},q$) and ($\log{\Delta},M_{tot,\sun}$) plots.  As
discussed in \citet{re01a}, it is reasonable that wide binaries
will have small $q$ (where $q \equiv m_{sec}/m_{pri}$) since
system binding energies are proportional to $M_{tot}/q$. The three
binaries in this paper, all with $q<0.15$, follow this
expectation, as do all previously discovered wide separation
binaries with L dwarf companions.

\citet{re01a} also find a characteristic upper limit radius for
binary separation as a function of total mass in their
($\log{\Delta},M_{tot,\sun}$) plot of L dwarf and M dwarf binaries
in the solar neighborhood and HST Hyades and field binaries, where
$\Delta$ has units of AU.  Empirically they find a cut-off defined
by the log-normal relation
\begin{equation}\label{binding}
\log{\Delta_{max}}=3.33M_{tot,\sun}+1.1
\end{equation}
The systems Gl 337 and HD 89744 have total masses over $1.5
M_\sun$ so their separations are at least three orders of
magnitude smaller than the maxima predicted by Equation
\ref{binding}. The Gl 618.1B system, an M-dwarf/L-dwarf pair with
$M_{tot,\sun} \sim 0.75$ and $\log{\Delta}=3.04$, comes closer to
the cut-off but is still less than the $\log{\Delta_{max}} \sim
3.43$ prediction from Equation \ref{binding}.  Thus all three
companions have $\Delta$ that fall well within the upper limits.

\section{Conclusion}
We have discovered two wide separation L-dwarf common proper
motion companions to nearby stars and identified a third candidate
from 2MASS. Spectral types assigned from optical spectroscopy were
L0 V, L2.5 V, and L8 V. NIR low resolution spectra of the
companions were provided as well as a grid of known objects
spanning M6 V -- T dwarfs to support spectral type assignment for
these and future L-dwarfs in the \textit{z'JHK} bands. Using
published measurements, we estimated ages of the companions from
physical properties of the primaries. These crude ages allowed us
to estimate companion masses using theoretical low-mass star and
brown dwarf evolutionary models.  We found that Gl 337C is a brown
dwarf, HD 89744B is a VLM star, and Gl 618.1B may be either. With
the addition of these new L-dwarfs there are nine wide-binary
($\Delta \geq 100$ AU) L-dwarf companions of nearby stars known.
These discoveries improve the statistics of binaries involving
low-mass stars and brown dwarfs. In particular they support
initial conclusions that the 'brown dwarf desert' seen at small
separations around main sequence stars does not extend to wide
separations.  The discovery of these companions in low mass ratio
systems ($q \la 0.12$), along with similar discoveries previously
reported, will also significantly improve our understanding of the
binary mass ratio distribution for $q<0.2$.

\acknowledgements We thank the anonymous referee for helpful
comments that led to a more concise manuscript, John Carr for a
careful reading of a previous draft, and Adam Burrows for
providing his evolutionary curves in electronic format.  We thank
Andrea Richichi and Isabelle Baraffe for supplying filter curves,
and Isabelle Baraffe and France Allard for helpful discussions. We
also wish to thank the Palomar Night Assistants Skip Staples and
Karl Dunscombe for their expertise and support. This publication
makes use of data from the Two Micron All Sky Survey, which is a
joint project of the University of Massachusetts and the Infrared
Processing and Analysis Center, funded by the National Aeronautics
and Space Administration and the National Science Foundation. This
research has also made use of the SIMBAD database, operated at
CDS, Strasbourg, France.  DSS images were obtained from the
Canadian Astronomy Data Centre, which is operated by the Herzberg
Institute of Astrophysics, National Research Council of Canada.

\clearpage

\clearpage

\begin{deluxetable}{lcccccc}
\tabletypesize{\scriptsize} \tablewidth{0pt}
\tablecaption{Candidate Companions \label{objects}}
\tablehead{\colhead{Object} & \colhead{RA (2000)} & \colhead{Dec
(2000)} & \colhead{$J$} & \colhead{$H$} & \colhead{$K_s$} &
\colhead{$J-K_s$}}

\startdata
2MASSW J0912145+145940 (2M0912+14) & 09:12:14.5 & +14:59:40 &
$15.70\pm0.08$ & $14.59\pm0.08$ & $14.02\pm0.06$ & $1.68\pm0.10$\\
2MASSW J1620261-041631 (2M1620-04) & 16:20:26.1 & -04:16:31 &
$15.31\pm0.05$ & $14.32\pm0.05$ & $13.59\pm0.04$ & $1.72\pm0.07$\\
2MASSI J1022148+411426 (2M1022+41) & 10:22:14.8 & +41:14:26 &
$14.89\pm0.04$ & $14.04\pm0.05$ & $13.62\pm0.05$ & $1.27\pm0.06$\\
\enddata
\end{deluxetable}

\clearpage

\begin{deluxetable}{lcccc}
\tabletypesize{\footnotesize} \tablewidth{0pt} \tablecaption{Log
of Spectroscopic Observations \label{observations}}
\tablehead{\colhead{Object} & \colhead{Obs Date (UT)} &
\colhead{Exposure (s)} & \colhead{Telescope} &
\colhead{Instrument}}

\startdata 2M0912+14 (Gl 337C) & 2000 May 10 & 4000 & Palomar
60-inch & CorMASS\\ \nodata & 2000 Dec 26 & 2400 & Keck I & LRIS\\
2M1620-04 (Gl 618.1B) & 2000 May 9 & 3300 & Palomar 60-inch &
CorMASS\\ \nodata & 2000 Oct 7 & 2400 & Keck I & LRIS\\ 2M1022+41
(HD 89744B) & 2000 May 10 & 4550 & Palomar 60-inch & CorMASS\\
\nodata & 2000 Dec 26 & 900 & Keck I & LRIS\\
\enddata
\end{deluxetable}

\clearpage

\begin{deluxetable}{llccccccc}
\tabletypesize{\scriptsize} \tablewidth{0pt} \tablecaption{Late-M
-- L-dwarf Grid Observations\label{grid log}}
\tablehead{\colhead{Sp T} & \colhead{Object} & \colhead{Obs Date
(UT)} & \colhead{Exposure (s)} & \colhead{$K_s$} & \colhead{$K$} &
\colhead{$J-K_s$} & \colhead{$J-K$} & \colhead{Ref}}

\startdata
M6 V & LHS 2034 & 1999 Oct 27 & 600 & \nodata & 10.03 & \nodata & 0.97 & 1\\
M8 V & VB 10 & 1999 Aug 22 & 720 & \nodata & 8.80 & \nodata & 1.10 & 1\\
M9 V & LHS 2924 & 2000 Jun 16 & 2400 & \nodata & 10.67 & \nodata & 1.17 & 1\\
M9.5 V & BRI 0021-0214 & 1999 Aug 24 & 1080 & \nodata & 10.64 & \nodata & 1.26 & 2\\
L1 V & 2MASSW J0208183+254253 & 1999 Aug 24 & 2700 & 12.58 & \nodata & 1.44 & \nodata & 3\\
L2 V & 2MASSW J0015447+351603 & 2000 Sep 18 & 3300 & 12.24 & \nodata & 1.58 & \nodata & 3\\
L3.5 V & 2MASSW J0036159+182110 & 1999 Aug 24 & 1440 & 11.03 & \nodata & 1.41 & \nodata & 4\\
L5 V & 2MASSW J1507476-162738 & 2000 Apr 20 & 2800 & 11.30 & \nodata & 1.52 & \nodata & 4\\
L7 V & DENIS-P J0205.4-1159AB & 2000 Sep 16 & 5100 & 12.99 & \nodata & 1.56 & \nodata & 5\\
L8 V & 2MASSW J1632291+190441 & 2000 Jun 14 & 5500 & 13.98 & \nodata & 1.88 & \nodata & 6\\
T & 2MASSW J0559191-140448 & 1999 Oct 24 & 2400 & 13.61 & \nodata & 0.22 & \nodata & 7\\
\enddata
\tablerefs{
(1) Leggett 1992; (2) Tinney et al. 1993; (3) Kirkpatrick et al. 2000;
(4) Reid et al. 2000; (5) Delfosse et al. 1997; (6) Kirkpatrick et al. 1999;
(7) Burgasser et al. 2000.}
\end{deluxetable}

\clearpage

\begin{deluxetable}{lcccccccc}
\tabletypesize{\scriptsize} \tablewidth{0pc}
\tablecaption{Compiled Data on Wide-Binary Primaries
\label{primproperties}} \tablehead{ \colhead{}
&\multicolumn{2}{c}{Gl 337AB} & \colhead{} & \multicolumn{2}{c}{Gl
618.1A} & \colhead{} & \multicolumn{2}{c}{HD 89744A} \\
\cline{2-3} \cline{5-6} \cline{8-9} \\ \colhead{} & \colhead{data}
& \colhead{ref} & \colhead{} & \colhead{data} & \colhead{ref} &
\colhead{} & \colhead{data} & \colhead{ref}}
\startdata Spectral Type & G8 V $+$ K1 V\tablenotemark{a} & 1 & & M0 V & 8 & & F7 IV-V & 11\\
\bv & 0.73 & 2 & & 1.4 & 9 & & 0.54 & 10 \\
Distance (pc) & $20.5\pm 0.4$ & 3 & & $30.3\pm 2.4$ & 3 &&
$39.0\pm 1.1$ & 3 \\
$M_{tot,\sun}$ & $1.74\pm0.15$ & 4 & & 0.67\tablenotemark{b} & - & & $1.41$\tablenotemark{b} & - \\
Activity & & & & & & & & \\
\phm{text}$\log{({L}_X/{L}_{bol})}$ & $-5.69$ & 5,6 & &
    not detected & 6 & & $<-5.7$ & 12 \\
\phm{text}$\log{(R^\prime_{HK})}$ & \nodata & & & \nodata & & &
    $-5.04$,$-5.12$ & 10,13 \\ \phm{text}$H\alpha$ (EW) & \nodata & &
    & $-0.62\pm0.35$ & 10 & & \nodata & \\
\phm{text}$P_{rot} \;(\rm{days})$ & \nodata & & & \nodata & & & 9 & 13 \\
\phm{text}period & \nodata & & & \nodata & & & flat & 14 \\
Metallicity & & & & & & & & \\ \phm{text}$\rm{[Fe/H]}$ &
$-0.14$\tablenotemark{c} & 2,7 & & \nodata & & &
    +0.24\tablenotemark{d} & 15,16 \\
\phm{text}$\rm{[M/H]}$ & \nodata & & & $\sim 0$ & 9 & & \nodata & \\
Lithium Abundance & & & & & & & & \\ \phm{text}$\log{N_{Li}}$ &
    \nodata & & & \nodata & & & 2.07\tablenotemark{e} & 16 \\
Kinematics & & & & & & & & \\ \phm{text}$U
(\rm{km/s})$\tablenotemark{f} & $-74$ & 7 & & $+122.3$ & 9 & & $+9.3$ & 15 \\
\phm{text}$V (\rm{km/s})$ & $-2$ & 7 & & $-62.2$ & 9 & & $-25.6$ & 15 \\
\phm{text}$W (\rm{km/s})$ & $+3$ & 7 & & $-12.8$ & 9 & & $-13.6$ & 15 \\
\enddata
\tablenotetext{a}{Combined spectral type from K0 V to G8 V have
been published for this object, with the latter favored by
\citet{mas96}. More recent measurements indicate G8 V + K1 V based
upon magnitude differences of the components \citep{bar00}.}
\tablenotetext{b}{See text \S \ref{primasses} for derivation.}
\tablenotetext{c}{The average of $\rm{P[Fe/H]}=-0.33$, derived
from Str\"{o}mgren photometry \citep{egg98} and
$\rm{P[Fe/H]}=0.04$, derived by \citet{arr89} using the
$(\rm{U-B})$ index of \citet{car79}.} \tablenotetext{d}{The
average of $\rm{[Fe/H]}=0.18$ \citep{edv93} and $\rm{[Fe/H]}=0.30$
\citep{gon01}.  Both measurements are from high resolution
spectroscopy.} \tablenotetext{e}{On the scale of [$\log{\epsilon
(Li_\sun)} \equiv \log{(N_{Li}/N_H)_\sun} + 12 = 1.06$].}
\tablenotetext{f}{$U$ defined positive towards the Galactic Center
(GC).  \citet{egg98}, \citet{leg92}, and \citet{edv93} define $U$
positive away from the GC, so their measurements have been
multiplied by $-1$.} \tablerefs{ (1) Barnaby et al. 2000; (2)
Arribas \& Roger 1989; (3) Perryman et al. 1997; (4) Pourbaix
2000; (5) Mason, McAlister, \& Hartkopf 1996; (6) H\"{u}nsch et
al. 1999; (7) Eggen 1998; (8) Reid, Hawley, \& Gizis 1995; (9)
Leggett 1992; (10) Soderblom 1985; (11) Saar \& Brandenburg 1999;
(12) Pizzolato, Maggio, \& Sciortino 2000; (13) Baliunas,
Sokoloff, \& Soon 1996; (14) Baliunas et al. 1995; (15) Edvardsson
et al. 1993; (16) Gonzalez et al. 2001.}
\end{deluxetable}

\clearpage

\begin{deluxetable}{lccc}
\tabletypesize{\footnotesize} \tablewidth{0pt}
\tablecaption{Derived Secondary Parameters \label{secparameters}}
\tablehead{\colhead{} & \colhead{Gl 337C} & \colhead{Gl 618.1B} &
\colhead{HD 89744B}}

\startdata Spectral Type  & L8 V & L2.5 V & L0 V \\
$T_{eff}$ (K) & $1300-1600$ & $1800-1950$ & $2000-2200$ \\
Age (Gyr) & $0.6-3.4$ & $0.5-12$ & $1.5-3.0$ \\
$M_{jup}$ & $40-74$ & $60-79$ & $77-80$ \\
\enddata
\end{deluxetable}

\clearpage

\begin{deluxetable}{lccc}
\tabletypesize{\footnotesize} \tablewidth{0pt}
\tablecaption{Wide-Binary System Parameters \label{sysparameters}}
\tablehead{\colhead{} & \colhead{Gl 337} & \colhead{Gl 618.1} &
\colhead{HD 89744}}

\startdata
$M_{tot,\sun}$ & $1.77-1.81$ & $0.73-0.75$ & $1.48$ \\
$q$ & $\la 0.04$ & $\la 0.12$ & $\la 0.06$ \\
$\Delta$ (AU) & 881 & 1090 & 2460 \\
\enddata
\end{deluxetable}

\end{document}